\documentclass[conference,a4paper]{IEEEtran}
\usepackage[dvipsnames]{xcolor}
\usepackage{balance}

\ifCLASSOPTIONcompsoc
 \usepackage[caption=false,font=normalsize,labelfont=sf,textfont=sf]{subfig}
\else
 \usepackage[caption=false,font=footnotesize]{subfig}
\fi
\usepackage[nolist,nohyperlinks]{acronym}
\usepackage{amsmath,amssymb}

\usepackage{graphicx}
\usepackage{epstopdf}
\usepackage{dblfloatfix}

\begin{document}
\title{Modeling Diffuse Scattering in Mobility Scenarios: Ray Tracing Model and Measurement Validation}

\author{\IEEEauthorblockN{
Milica Djordjevic,
Thomas Wilding,
Anja Dakić,
Benjamin Rainer,
Markus Hofer,
Thomas Zemen
}
\IEEEauthorblockA{
AIT Austrian Institute of Technology, Vienna, Austria\\
Email: milica.djordjevic@ait.ac.at
}
}



\newcommand{\com}[1]{{\color{blue}\textit{#1}}}
\newcommand{\md}[1]{{\color{red}md: #1}}
\newcommand{\tw}[1]{{\color{Cyan}tw: #1}}

\setlength{\abovedisplayskip}{5pt}
\setlength{\belowdisplayskip}{5pt}
\setlength{\abovedisplayshortskip}{3pt}
\setlength{\belowdisplayshortskip}{3pt}
\maketitle

\begin{abstract}
Open-source ray tracing (RT) frameworks are increasingly used for site-specific channel modeling due to their reproducibility and extensibility. In non-stationary mobility scenarios, repeated RT evaluations can generate different diffuse propagation paths without preserving their correspondence, limiting a consistent representation of their temporal evolution. To address this limitation, we extend diffuse scattering modeling in Sionna RT using persistent surface-associated scattering points. Their positions remain fixed, while their contributions evolve according to changes in the propagation geometry and visibility. The proposed approach is validated using empirical radio channel impulse response measurements in an urban vehicular scenario at 5.9~GHz. Compared with the standard Sionna RT baseline, the proposed model shows improved agreement with measurements.
\end{abstract}

\vskip0.5\baselineskip

% \fi
% \IEEEpeerreviewmaketitle
\begin{acronym} 
\setlength{\itemsep}{-0.63\parsep}
\acro{3GPP}{3rd Generation Partnership Project}
\acro{ADAS}{advanced driver assistance system}
\acro{AoA}{angle of arrival}
%\acro{AOD}{angle of departure}
\acro{ACF}{auto correlation function}
%\acro{ASIC}{application-specific integrated circuit}
\acro{ADC}{analog-to-digital converter}
\acro{APP}{a posteriori probability}
%\acro{AWGN}{additive white Gaussian noise}
\acro{BEM}{basis-expansion model}
\acro{BER}{bit error rate}
\acro{BLER}{block error rate}
\acro{BS}{base station}
%\acro{BER}{bit-error-rate}
%\acro{CA}{complex addition}
\acro{CE}{complex exponential}
\acro{CIR}{channel impulse response}
\acro{CFR}{channel frequency response}
%\acro{CDMA}{code division multiple access}
%\acro{CM}{complex multiplication}
%\acro{COTS}{common-off-the-shelf}
\acro{CRC}{cyclic redundancy check}
\acro{CP}[CP]{cyclic prefix}
\acro{CCAM}{cooperative connected automated mobility}
\acro{CDF}{cumulative distribution function}
%\acro{CSM}{complex scalar multiplication}
\acro{CTF}{channel transfer function}
\acro{CSI}{channel state information}
\acro{D2D}{device-to-device}
\acro{DAC}{digital-to-analog converter}
\acro{DC}{direct current}
%\acro{DDC}{direct down conversion}
%\acro{DMA}{direct memory access}
\acro{DOD}{direction of departure}
\acro{DOA}{direction of arrival}
%\acro{DUC}{direct up conversion}
\acro{DPS}{discrete prolate spheroidal}
\acro{DPSWF}{discrete prolate spheroidal wave function}
\acro{DSD}{Doppler spectral density}
\acro{DSP}{digital signal processor}
\acro{DR}{dynamic range}
%\acro{eNodeB}{base station}
%\acro{eMBB}{enhanced mobile broadband} %used
\acro{ETSI}{European Telecommunications Standards Institute}
%\acro{FER}{frame error rate}
\acro{FRMCS}{future railway mobile communication system}
\acro{FIFO}{first-input first-output}
\acro{FFT}{fast Fourier transform}
%\acro{FLOPS}{floating point operations per second}
\acro{FP}{fixed-point}
\acro{FPGA}{field programmable gate array}
%\acro{FPS}{frames per second}
%\acro{GCE}{geometry-based channel emulator}
\acro{GSCM}{geometry-based stochastic channel model}
%\acro{GCM}{geometry-based channel model}
\acro{GSM}{global system for mobile communications}
\acro{GPS}{global positioning system}
%\acro{HiL}{hardware-in-the-loop}
\acro{HPBW}{half power beam width}
\acro{ICI}[ICI]{inter-carrier interference}
\acro{IDFT}{inverse discrete Fourier transform}
\acro{IF}{intermediate frequency}
\acro{IFFT}{inverse fast Fourier transform}
\acro{IMU}{inertial measurement unit}
\acro{ISI}{inter-symbol interference}
\acro{ITS}{intelligent transportation system}
\acro{IXL}{interlocking}
%\acro{mMTC}{massive machine type communications} %used
%\acro{MAC}{multiply-accumulate}
\acro{JSAC}{joint sensing and communication}
\acro{JCAS}{joint communication and sensing}
\acro{MAC}{medium access control}
\acro{MEC}{mobile edge computing}
\acro{MSE}{mean square error}
\acro{LLR}{log-likelihood ratio}
\acro{LO}{local oscillator}
\acro{LOS}{line-of-sight}
\acro{LMMSE}{linear minimum mean squared error}
\acro{LNA}{low noise amplifier}
\acro{LS}{least squares}
\acro{LSF}{local scattering function}
\acro{LTE}{long term evolution}
%\acro{LTE-A}{Long Term Evolution Advanced}
\acro{LUT}{look-up table}
\acro{LTV}{linear time-variant}
\acro{LX}{level crossing}
\acro{MIMO}{multiple-input multiple-output}
\acro{MMSE}{minimum mean squared error}
\acro{MMW}[mmWave]{millimeter wave}
\acro{MPC}{multipath component}
\acro{MUSIC}{multiple signal classification}
\acro{MC}{Monte Carlo}
\acro{NI}{National Instruments}
\acro{NLOS}{non-line of sight}
\acro{OC}{object controller}
\acro{OFDM}{orthogonal frequency division multiplexing}
\acro{OLOS}{obstructed line-of-sight}
\acro{OSM}{OpenStreetMap}
\acro{OTA}{over-the-air}
\acro{PA}{power amplifier}
\acro{PC}{personal computer}
\acro{PDP}{power delay profile}
%\acro{PDF}{probability density function}
\acro{PER}{packet error rate}
\acro{PPS}{pulse per second}
%\acro{P2P}{peer-to-peer}
\acro{QAM}{quadrature ampltiude modulation}
\acro{QPSK}{quadrature phase shift keying}
%\acro{UE}{user equipment}
\acro{RaSTA}{rail safe transport application}
\acro{RB}{resource block}
\acro{RBP}{resource block pair}
\acro{RF}{radio frequency}
\acro{RMS}{root mean square}
\acro{RSSI}{receive signal strength indicator}
\acro{RT}{ray tracing}
\acro{RTK}{real time kinematic}
\acro{RX}{receiver}
\acro{SBR}{shooting-and-bouncing rays}
\acro{SCME}{spatial channel model extended}
\acro{SDR}{software defined radio}
\acro{SIMO}{single-input multiple-output}
\acro{SISO}{single-input single-output}
\acro{SoCE}{sum of complex exponentials}
\acro{SNR}{signal-to-noise ratio}
\acro{SUT}{system-under test}
\acro{SSD}{soft sphere decoder}
\acro{TB}{transport block}
\acro{TBWP}{time-bandwidth product}
\acro{TDL}{tap delay line}
%\acro{TTI}{transmission time intervall}
\acro{TX}{transmitter}
%\acro{UAV}{unmanned ariel vehicle}
\acro{UMTS}{universal mobile telecommunications systems}
\acro{UDP}{user datagram protocol}
\acro{URLLC}{ultra-reliable and low latency communication} %yes
\acro{US}{uncorrelated-scattering}
\acro{USRP}{universal software radio peripheral}
\acro{UTM}{universal transverse mercator}
\acro{VAA}{virtual antenna array}
\acro{VLA}{virtual linear array}
\acro{VNA}{vector network analyzer}
\acro{ViL}{vehicle-in-the-loop}
\acro{V2I}{vehicle-to-infrastructure}
\acro{V2V}{vehicle-to-vehicle}
\acro{V2X}{vehicle-to-everything}
\acro{VST}{vector signal transceiver}
\acro{VTD}{Virtual Test Drive}
\acro{WF}{Wiener filter}
\acro{WSS}{wide-sense-stationary}
\acro{WSSUS}{wide-sense-stationary uncorrelated-scattering}
\end{acronym}
\section{Introduction}
Reliable site-specific channel simulations are important for the design and evaluation of modern wireless communication systems, particularly in dynamic urban and vehicular environments. \Ac{RT} provides a geometry-based approach for modeling multipath propagation, including \ac{LOS} propagation, specular reflection, diffraction, and diffuse scattering \cite{Zemen2025SSCR,Fuschini2015RTReview}. In vehicular scenarios, mobility continuously changes the propagation geometry, resulting in a non-stationary channel and a time-varying \ac{CIR}. Its temporal evolution can be characterized in the delay and Doppler domains using the \ac{PDP}, \ac{DSD}, and the corresponding \ac{RMS} delay and Doppler spreads \cite{bernado2014delay}. An accurate representation of this evolution requires modeling not only the dominant propagation paths but also the temporal evolution and correlation of diffuse multipath components.

Different approaches have been proposed to model diffuse multipath components in mobile propagation scenarios. Geometry-based stochastic models consider scatterers distributed along roads in vehicular environments \cite{KaredalTWC2009} or within geometry-dependent clusters in street canyons \cite{HuangTWC2021}. In \cite{SchubertTAP2013}, a non-stationary land-mobile satellite channel model represents scattering from vegetation and is validated against measurements. In \ac{RT}-based channel modeling, diffuse scattering has been incorporated through surface-based approaches such as the Effective Roughness model, where scattering surfaces are discretized into tiles acting as secondary diffuse sources \cite{DegliEsposti2007,Vitucci2012}. In \cite{RainerPIMRC2020}, a position-invariant lattice-based surface representation was introduced to obtain spatially correlated diffuse scattering and a time-varying \ac{DSD}.

Sionna RT provides an open-source and GPU-accelerated framework for site-specific radio propagation modeling \cite{Hoydis2023SionnaRT,AitAoudia2025SionnaRT}. However, when propagation paths are recomputed at successive channel snapshots, its standard diffuse scattering implementation does not explicitly preserve the identity of diffuse scattering points across \ac{RT} evaluations. This can introduce discontinuities in the temporal evolution of diffuse components, particularly in mobility scenarios where the propagation geometry changes continuously. To address this limitation, we propose an extension of diffuse scattering modeling within Sionna RT based on persistent surface-associated scattering points. The scattering points remain fixed with respect to the scene geometry, while their visibility and contributions evolve with transmitter and receiver motion. This enables a consistent temporal representation of diffuse multipath components while retaining the standard Sionna RT models for \ac{LOS}, specular reflection, and diffraction. Since the formulation is based on the scene geometry, it can be applied to different mobility scenarios, including urban, rural, maritime, and satellite-related environments.

\textit{Scientific Contributions of the Paper:}
\begin{itemize}
    \item An extension of diffuse scattering modeling in Sionna RT using persistent surface-associated scattering points to preserve the identity of diffuse components across channel snapshots.

    \item A geometry-dependent visibility and activation mechanism that gradually adjusts the contributions of scattering points as the transmitter and receiver move.

    \item Validation using measured vehicular \acp{CIR} at $5.9\,\mathrm{GHz}$, comparing the proposed approach with the standard Sionna RT model in terms of \ac{PDP}, \ac{DSD}, \ac{RMS} delay spread, and \ac{RMS} Doppler spread.
\end{itemize}

\section{Diffuse scattering}
\subsection{Propagation Mechanisms and Channel Model}
The considered channel model includes \ac{LOS} propagation, specular reflection, diffraction, and diffuse scattering. Specular reflection follows the direction defined by geometrical optics, while diffuse scattering accounts for non-specular propagation caused by surface roughness, structural details, and material inhomogeneities \cite{DegliEsposti2007,Vitucci2012}. The scattered energy is distributed over a wider range of propagation directions, described by a scattering pattern. In dynamic scenarios, mobility changes the propagation geometry and the corresponding propagation paths, resulting in time-varying path gains, delays, and Doppler shifts. The time-varying channel frequency response is expressed as
\begin{align}
H(t,f) = \sum_{p=1}^{P(t)} \alpha_p(t)e^{-\iota2\pi f \tau_p(t)} =
H_{\mathrm{d}}(t,f) + H_{\mathrm{s}}(t,f),
\label{eq:time_varying_cfr}
\end{align}
where $p$ denotes the propagation path index, $P(t) \in \mathbb{N}$ is the number of propagation paths at time $t$, and $\alpha_p(t) \in \mathbb{C}$ and $\tau_p(t) \in \mathbb{R}^{+}$ denote the time-varying complex path gain and propagation delay of path $p$, respectively. The terms $H_{\mathrm{d}}(t,f)$ and $H_{\mathrm{s}}(t,f)$ denote the deterministic and diffuse scattering contributions, respectively. The temporal evolution of the propagation paths determines the non-stationary behavior of the channel. In this work, the main focus is on the temporal evolution of the diffuse scattering contribution $H_{\mathrm{s}}(t,f)$.

\subsection{Sionna RT Baseline}
\begin{figure}[!t]
    \centering
    \includegraphics[width=0.85\columnwidth]{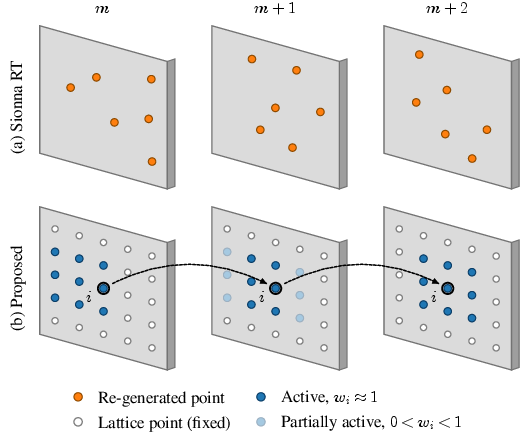}
    \caption{Diffuse scattering evolution across three snapshots. (a) Sionna RT generates diffuse interaction points without preserving their correspondence. (b) The proposed model preserves fixed scattering points with time-varying activation weights. Arrows indicate the preserved identity of point $i$.}
    \label{fig:diffuse}
\end{figure}

In Sionna RT, diffuse scattering is modeled using material-dependent scattering properties. The scattered power is controlled by the scattering coefficient $S$, while the angular distribution is described by Lambertian, directive, or backscattering patterns based on~\cite{DegliEsposti2007}. Diffuse paths are generated within the \ac{SBR} procedure by sampling the interaction type and outgoing diffuse direction. For the baseline considered in this work, \ac{LOS}, specular reflection, diffraction, and diffuse scattering are computed using the standard Sionna RT propagation models. For mobility, Sionna RT can evolve previously computed paths based on their Doppler shifts. This approach is intended for short trajectories, typically not exceeding a few wavelengths~\cite{AitAoudia2025SionnaRT}. For longer trajectories, the propagation paths need to be recomputed as the transmitter and receiver move. The diffuse paths are therefore generated by the \ac{SBR} procedure at successive \ac{RT} evaluations, without persistent correspondence between them. Consequently, the baseline model does not preserve the temporal evolution of individual diffuse scattering components. This limitation is illustrated in Fig.~\ref{fig:diffuse} and motivates the proposed extension.

\subsection{Proposed Extension of Sionna RT}
To provide a temporally correlated evolution of diffuse multipath components in mobility scenarios, we extend the diffuse scattering model in Sionna RT version 2.2.0~\cite{sionna} using persistent surface-associated scattering points. Following the position-invariant surface representation introduced in~\cite{RainerPIMRC2020}, each point $i$ is generated once and characterized by its fixed position $\mathbf{p}_i$, surface normal $\mathbf{n}_i$, effective area $\Delta E_i$, and random phase $\phi_i\sim\mathcal{U}(0,2\pi)$. These parameters remain fixed throughout the simulation, while the scattering contributions evolve with mobility.

Before the simulation, triangles are grouped into coplanar surface regions. A local two-dimensional orthonormal basis is constructed using the Gram-Schmidt procedure to generate a regular lattice with spacing $\Delta_s=c/(2B)$, where $c$ is the speed of light in vacuum and $B$ is the considered bandwidth. The lattice basis vectors are defined as~\cite{RainerPIMRC2020}
\begin{equation}
    \mathbf{b}_i = \mathbf{o}_i \Delta_s,
    \qquad i\in\{1,2\},
    \label{eq:lattice_basis}
\end{equation}
where $\mathbf{o}_i$ denotes the corresponding orthonormal basis vector. Only points within the corresponding surface regions are retained. For a region $r$ with area $E_r$ and $N_r$ scattering points, each point is assigned an effective area $\Delta E_i=E_r/N_r$ before visibility filtering. This preserves the represented surface area of the initial lattice, while the assigned effective areas remain unchanged after filtering.

The initial lattice is restricted to the region surrounding the transmitter and receiver trajectories. A visibility prescan at selected trajectory positions retains points simultaneously visible from the transmitter and receiver at least once, forming the set $\mathcal{I}$. During the simulation, the instantaneous visibility $\chi_i[m]\in\{0,1\}$ is evaluated at each snapshot $m$. To avoid abrupt changes in diffuse contributions, we propose an activation-weight model in which the weight $w_i[m]\in[0,1]$ is updated as
\begin{equation}
    w_i[m]=\alpha_m w_i[m-1]+(1-\alpha_m)\chi_i[m],
    \label{eq:scatterer_weight}
\end{equation}
where
\begin{equation}
    \alpha_m=\exp\left(
    -\frac{v_{\mathrm{eff}}[m]\Delta t}{L_{\mathrm f}}
    \right),\qquad L_{\mathrm f}=\Delta_s.
    \label{eq:fade_factor}
\end{equation}
Here, $\Delta t$ is the snapshot interval and $v_{\mathrm{eff}}[m]$ denotes the effective velocity calculated from the transmitter and receiver motion. As visibility changes, scattering contributions gradually activate or deactivate. Path lengths and interaction angles are recomputed from the current geometry, while the activation weight retains information from the previous snapshot. Contributions are included only for geometrically valid interactions with activation weights above the selected threshold.

The implementation considers first-order diffuse scattering without interactions between diffuse scattering points. Following the directive scattering formulation in~\cite{RainerPIMRC2020}, the amplitude coefficient is
\begin{equation}
    A_i[m]=\frac{SU}{d_i[m]}
    \sqrt{\frac{\Delta E_i\cos\theta_{\mathrm{t},i}[m]}
    {F(\alpha_R)}}
    \left(\frac{1+\cos\psi_i[m]}{2}\right)^{\alpha_R/2},
    \label{eq:diffuse_amplitude}
\end{equation}
where $S$ is the scattering coefficient, $U$ is the ratio of reflected to incident electric field amplitudes, which can be obtained from the Fresnel equations, and $d_i[m]$ denotes the total propagation length. The angle $\psi_i[m]$ is defined between the scattered and specular reflection directions, $\alpha_R=2$ is selected for the scattering lobe, consistent with~\cite{Xia2024}, and $F(\alpha_R)$ is the normalization factor defined in~\cite{RainerPIMRC2020}. The corresponding complex coefficient and propagation delay are
\begin{equation}
    a_i[m]=w_i[m]A_i[m]e^{j\phi_i},
    \qquad \tau_i[m]=\frac{d_i[m]}{c}.
    \label{eq:diffuse_coefficient}
\end{equation}

The diffuse \ac{CFR} is obtained by coherently summing the contributions of the scattering points retained after the visibility prescan,
\begin{equation}
    H_{\mathrm{s}}[m,q]=
    \sum_{i\in\mathcal{I}}a_i[m]
    e^{-\iota 2\pi f_q\tau_i[m]},
    \label{eq:diffuse_cfr}
\end{equation}
where $f_q$ denotes the $q$th frequency sample within the considered bandwidth. For points that do not satisfy the geometric validity and activation conditions, the contribution is set to zero.

For the proposed model, the built-in Sionna RT diffuse scattering mechanism is disabled. The \ac{LOS}, specular reflection, and diffraction contributions are obtained using the standard Sionna RT models. Consistent with~\eqref{eq:time_varying_cfr}, the total channel is written as
\begin{equation}
    H[m,q]=H_{\mathrm{d}}[m,q]+H_{\mathrm{s}}[m,q].
    \label{eq:total_cfr}
\end{equation}
The corresponding band-limited \ac{CIR} is obtained by applying an inverse fast Fourier transform along the frequency axis.
\section{Channel Model Validation}
\subsection{Measurement Scenario and Site-Specific Reconstruction}
The proposed model is evaluated using a vehicular measurement campaign conducted at an urban intersection in Kaiserm\"uhlen, Vienna, with two moving cars acting as transmitter and receiver~\cite{RainerPIMRC2020}. The measurements were recorded with the AIT multi-node channel sounder~\cite{Zelenbaba2020Sounder} operated at $5.9\,\mathrm{GHz}$ with a bandwidth of $150\,\mathrm{MHz}$, a frequency spacing of $250\,\mathrm{kHz}$, and a snapshot interval of $0.5\,\mathrm{ms}$. The transmitter and receiver antennas were mounted on the roofs of the corresponding vehicles at heights of approximately $1.4\,\mathrm{m}$ and $2\,\mathrm{m}$, respectively. Half-wavelength dipole antennas were used.

\begin{figure}[t]
    \centering
    \includegraphics[width=\columnwidth]{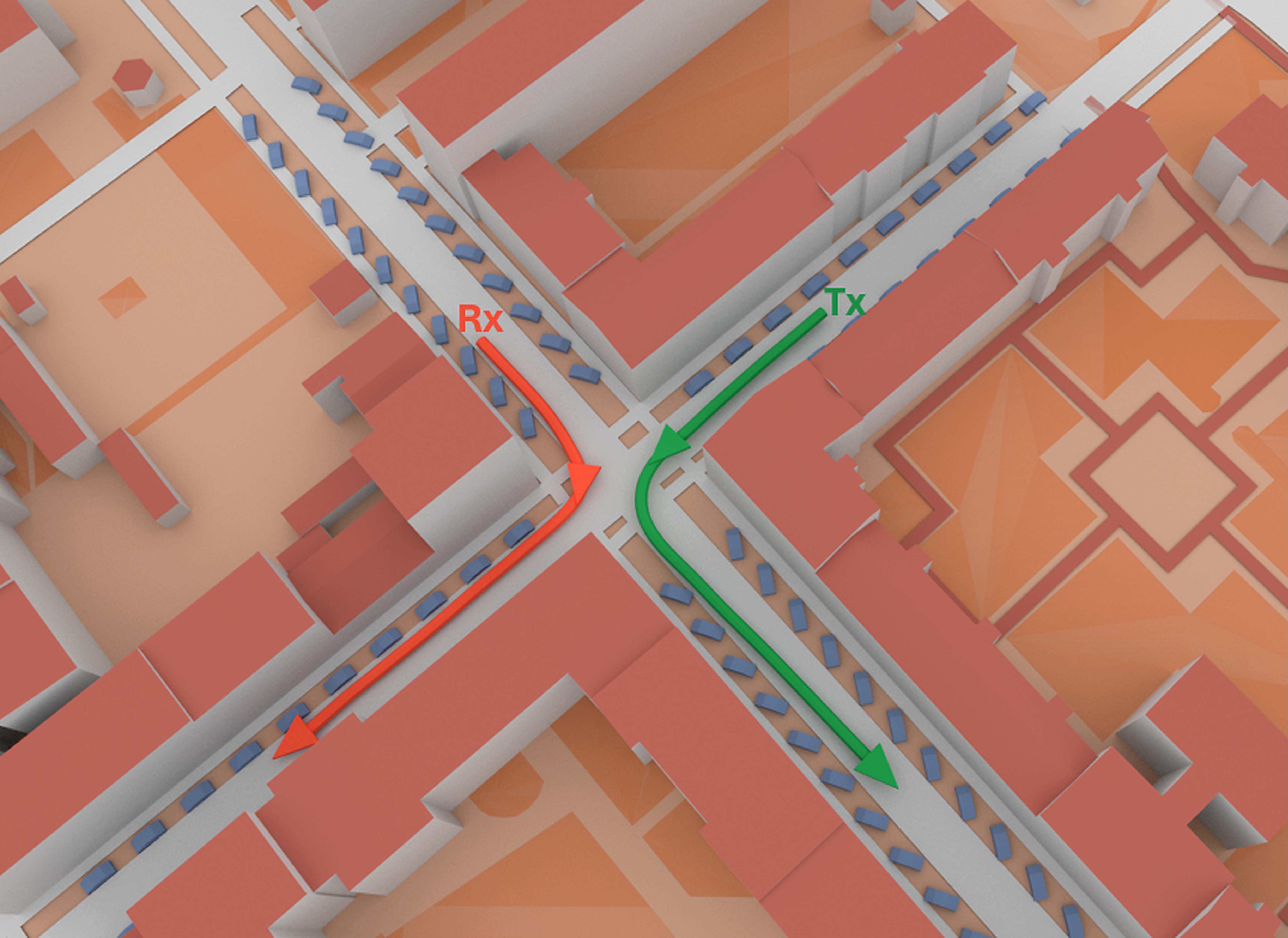}%\vspace{-2mm}
    \caption{Site-specific Kaiserm\"uhlen scene in Sionna RT with the Tx and Rx trajectories, parked vehicles, and the corresponding directions of motion.}
    \label{fig:scene}
\end{figure}

For the RT simulations, the measurement environment was reconstructed using data from OpenStreetMap, as shown in Fig.~\ref{fig:scene}. The building and street geometry was prepared in Blender and exported in Mitsuba XML format for use in Sionna RT. The recorded transmitter and receiver position samples were mapped to the reconstructed scene and spline-interpolated onto the simulation time grid. The scene also includes the parked vehicles that were present during the measurements.
%\begin{figure}[ht!]
    %\centering
    %\includegraphics[width=\columnwidth]{pics/Kaisermuhlen.jpg}
    %\caption{3D view of the considered environment in Kaiserm\"uhlen (©Google).}
    %\label{fig:Kaisermuhlen}
%\end{figure}
\subsection{Channel Evaluation Metrics}
\begin{figure*}[!b]
    \centering
    \includegraphics[width=0.98\textwidth]{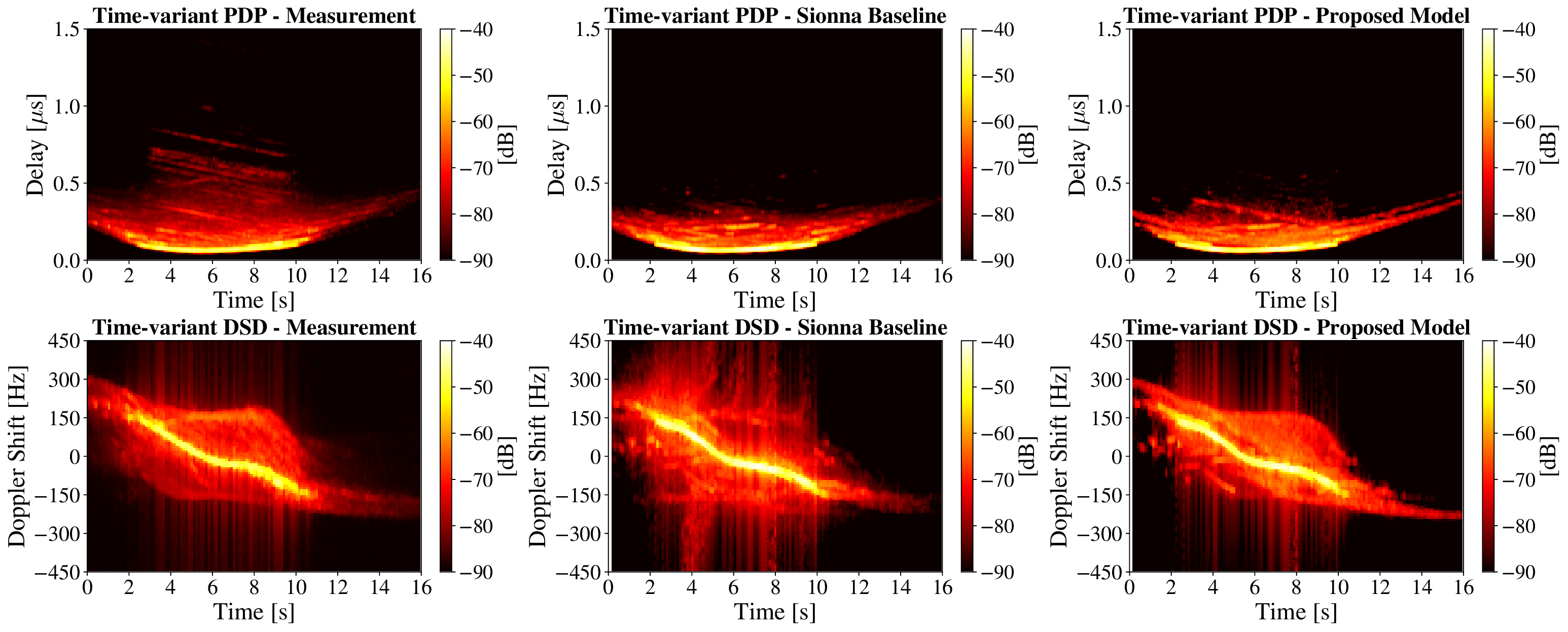}
    \caption{Time-variant PDP and DSD for the measurement, the Sionna RT baseline, and the proposed diffuse scattering model.}
    \label{fig:pdp_dsd}
\end{figure*}
The measured and simulated channels are processed using the same
procedure within finite stationarity regions. The sampled \ac{CFR}
$H[m,q]=H(m t_s,q f_s)$ is evaluated at discrete time and frequency
indices $m$ and $q$, with sampling intervals $t_s$ and $f_s$,
respectively. For each stationarity region $l$ with $M$ time and $N$ frequency samples, the \ac{LSF} is estimated using a multi-taper
approach~\cite{Bernado2009}
\begin{equation}
\widehat{C}[l;n,k]
=
\frac{1}{IJ}
\sum_{\rho=0}^{IJ-1}
\left|
H^{(G_\rho)}[l;n,k]
\right|^2 ,
\label{eq:lsf}
\end{equation}
where $n\in\{0,\ldots,N-1\}$ and $k\in\{-M/2,\ldots,M/2-1\}$ are the delay and Doppler indices, and $I$ and $J$ denote the numbers of time- and frequency-domain tapers. The tapered CFR is
\begin{align}
H^{(G_\rho)}[l;n,p]
&=
\sum_{m=-M/2}^{M/2-1}
\sum_{q=0}^{N-1}
H[m+Ml,q] \nonumber\\
&\quad {}\times
G_{iJ+j}[m,q]\,
e^{-\iota 2\pi(pm-nq)} .
\label{eq:tapered_cfr}
\end{align}
where the two-dimensional taper is formed from time- and frequency-domain discrete prolate spheroidal sequences (DPSS)~\cite{Slepian1978, bernado2014delay}, $G_{iJ+j}[m,q]=u_i[m]\,v_j[q]$. Here, $u_i[m]$ and $v_j[q]$ denote the corresponding time- and frequency-domain DPSS tapers, respectively.

The time-variant \ac{PDP} and \ac{DSD} are obtained as the marginals of the LSF~\cite{bernado2014delay} and are evaluated for each stationarity region, providing a local characterization of the time-varying channel in the delay and Doppler domains,
\begin{equation}
\begin{aligned}
\widehat{P}_{\tau}[l;n]
&=
\frac{1}{M}
\sum_{k=-M/2}^{M/2-1}
\widehat{C}[l;n,k],\, \widehat{p}_\tau[l;n] = \frac{\widehat{P}_{\tau}[l;n]}{\sum_{n = 0}^{N-1} \widehat{P}_{\tau}[l;n]},\\
\widehat{P}_{\nu}[l;k]
&=
\frac{1}{N}
\sum_{n=0}^{N-1}
\widehat{C}[l;n,k],\, \widehat{p}_\nu[l;k] = \frac{\widehat{P}_{\nu}[l;k]}{\sum_{k = M/2}^{M/2-1} \widehat{P}_{\nu}[l;k]}.
\end{aligned}
\label{eq:pdp_dsd}
\end{equation}
The delay and Doppler sampling intervals are $\tau_s = \frac{1}{Nf_s}$ and $\nu_s = \frac{1}{Mt_s}$. From the discrete densities one computes the mean delay, \ac{RMS} delay spread, mean Doppler frequency and \ac{RMS} Doppler spreads for each stationarity index $l$ by
\begin{equation}
    \begin{aligned}
        \overline{\tau}[l] &= \mathbb{E}_{\widehat{p}_\tau[l;\cdot]}[\tau_s],\, \sigma_{\tau}[l]^2 = \mathbb{V}_{\widehat{p}_\tau[l;\cdot]}[\tau_s], \\
        \overline{\nu}[l] &= \mathbb{E}_{\widehat{p}_\nu[l;\cdot]}[\nu_s], \, \sigma_{\nu}[l]^2 = \mathbb{V}_{\widehat{p}_\nu[l;\cdot]}[\nu_s].
    \end{aligned}
\end{equation}
%

% \begin{equation}
% \overline{\tau}[l]
% =
% \frac{
% \sum_{n=0}^{N-1}
% (n\tau_s)\widehat{P}'_{\tau}[l;n]
% }{
% \sum_{n=0}^{N-1}
% \widehat{P}'_{\tau}[l;n]
% },
% \label{eq:mean_delay}
% \end{equation}
% and
% \begin{equation}
% \sigma_{\tau}[l]
% =
% \sqrt{
% \frac{
% \sum_{n=0}^{N-1}
% (n\tau_s)^2\widehat{P}_{\tau}[l;n]
% }{
% \sum_{n=0}^{N-1}
% \widehat{P}_{\tau}[l;n]
% }
% -
% \overline{\tau}^{\,2}[l]
% }.
% \label{eq:delay_spread}
% \end{equation}
% %
% The corresponding mean Doppler frequency and \ac{RMS} Doppler spread are
% %
% \begin{equation}
% \overline{\nu}[l]
% =
% \frac{
% \sum_{k=-M/2}^{M/2-1}
% (k\nu_s)\widehat{P}'_{\nu}[l;k]
% }{
% \sum_{k=-M/2}^{M/2-1}
% \widehat{P}'_{\nu}[l;k]
% },
% \label{eq:mean_doppler}
% \end{equation}
% %
% and
% %
% \begin{equation}
% \sigma_{\nu}[l]
% =
% \sqrt{
% \frac{
% \sum_{k=-M/2}^{M/2-1}
% (k\nu_s)^2\widehat{P}'_{\nu}[l;k]
% }{
% \sum_{k=-M/2}^{M/2-1}
% \widehat{P}'_{\nu}[l;k]
% }
% -
% \overline{\nu}^{\,2}[l]
% }.
% \label{eq:doppler_spread}
% \end{equation}

\subsection{Validation Results}

The same scene geometry, transmitter and receiver trajectories, and material properties are used for both the Sionna RT baseline and the proposed model, such that only the diffuse scattering model differs. The reconstructed scene includes buildings and parked vehicles, whose surfaces are considered in the diffuse scattering calculation. For the Sionna RT baseline, $10^6$ samples per source, a maximum propagation depth of five interactions, and a limit of $5\times10^5$ paths per source are used. The number of generated diffuse paths depends on the ray sampling budget and propagation geometry, reaching approximately $4\times10^5$ at $t\approx5\,\mathrm{s}$, close to the vehicles' closest approach at approximately $5.3\,\mathrm{s}$. In the proposed model, the scattering-point lattice is generated on the reconstructed surfaces and restricted to the region surrounding the transmitter and receiver trajectories. For $B=150\,\mathrm{MHz}$, the lattice spacing is $\Delta_s=c/(2B)\approx1\,\mathrm{m}$. After the trajectory-dependent visibility prescan, 7621 scattering points are retained. The number simultaneously visible from the transmitter and receiver varies along the trajectory, reaching approximately 73\% of the retained points around the closest approach and decreasing to approximately 3\% towards the end. Unlike the baseline, the number of persistent scattering points is determined by the lattice spacing, scene geometry, and visibility prescan rather than the ray sampling budget.

The measured and simulated \acp{CIR} are processed using the same evaluation procedure. The measured transmitter and receiver position samples are spline-interpolated and evaluated on the simulation time grid with $t_s=0.5\,\mathrm{ms}$. A stationarity interval of $T_{\mathrm{stat}}=120\,\mathrm{ms}$ is used, corresponding to $M=240$ time samples per stationarity region.

The comparison of the measured and simulated time-variant PDP and DSD is shown in Fig.~\ref{fig:pdp_dsd}. Both simulated channels reproduce the dominant propagation components, while differences remain in the diffuse part of the channel. In the delay domain, both models mainly capture the dominant components, while the measurement contains additional weaker contributions over a wider delay range. Although the proposed model introduces additional diffuse components, the improvement in the PDP remains limited.

A more pronounced improvement is observed in the Doppler domain. While both models reproduce the dominant Doppler trajectory, the Sionna RT baseline mainly consists of distinct components and does not reproduce the continuous diffuse Doppler contribution observed in the measurement. In contrast, the proposed model produces a broader and more continuous diffuse Doppler distribution around the dominant trajectory, more closely resembling the measured diffuse cloud. Since the scattering points are preserved along the trajectory, their contributions evolve according to changes in the propagation geometry and visibility, providing a more consistent temporal evolution of diffuse scattering.

\begin{figure}[h!]
    \centering
    \includegraphics[width=\columnwidth]{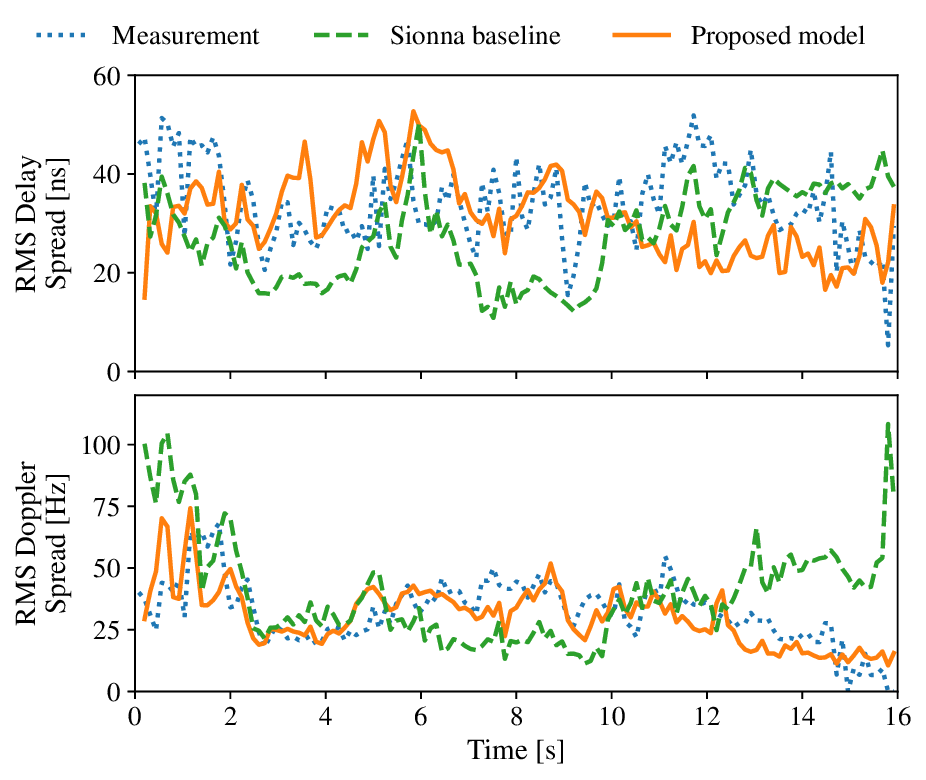}
    \caption{Comparison of RMS delay spread and RMS Doppler spread obtained from the measurement, the Sionna RT baseline, and the proposed diffuse scattering model.}
    \label{fig:rms}
\end{figure}

For the RMS delay and Doppler spreads, a $5\,\mathrm{dB}$ margin above the estimated noise floor and a $35\,\mathrm{dB}$ dynamic-range threshold below the respective peak are applied. The results are shown in Fig.~\ref{fig:rms}. In the delay domain, the proposed model provides only limited improvement compared with the baseline, and differences from the measured RMS delay spread remain. In contrast, the proposed model follows the measured RMS Doppler spread more closely, particularly between approximately $2$ and $12\,\mathrm{s}$. The temporal evolution of the RMS Doppler spread shows better agreement with the measurement than the baseline. The remaining discrepancies between $12$ and $16\,\mathrm{s}$ may be related to higher-order diffuse scattering interactions, which are not considered in the current implementation and will be investigated in future work. Overall, the main improvement is observed in the Doppler domain, where preserving the temporal evolution of diffuse components provides a more realistic representation of the time-varying channel.

\section{Conclusion}
In this paper, we extend the diffuse scattering model in Sionna RT for site-specific simulations of dynamic propagation scenarios. The proposed approach keeps surface-associated scattering points over successive channel snapshots, which allows the diffuse multipath components to evolve continuously along the transmitter and receiver trajectories. The model is validated using urban vehicular measurements at $5.9\,\mathrm{GHz}$ and evaluated in terms of \ac{PDP}, \ac{DSD}, \ac{RMS} delay spread, and \ac{RMS} Doppler spread. Compared with the Sionna RT baseline, the proposed model provides a more consistent temporal evolution of diffuse scattering components and better reproduces the measured Doppler characteristics, particularly the diffuse Doppler distribution and RMS Doppler spread, while improvements in the delay domain remain limited.

The results show the importance of preserving the temporal evolution of diffuse scattering components in mobility scenarios. Future work will investigate higher-order diffuse scattering interactions, different scattering point distributions, and the applicability of the proposed model to other propagation scenarios.
% use section for acknowledgement
\section*{ACKNOWLEDGEMENT}
This work has received funding from the European Defence Fund (EDF) under the grant agreement FESPAN no. 101167641. Funded by the European Union. Views and opinions expressed are those of the author(s) and do not necessarily reflect those of the European Union or the European Commission. Neither the European Union nor the granting authority can be held responsible for them.

% \IEEEtriggeratref{7}
% \IEEEtriggercmd{\enlargethispage{-20cm}}

\bibliographystyle{IEEEtran}
\bibliography{references}

@inproceedings{RainerPIMRC2020,
  author    = {Rainer, Benjamin and L{\"o}schenbrand, David and
               Zelenbaba, Stefan and Hofer, Markus and Zemen, Thomas},
  title     = {Towards a non-stationary correlated fading process for
               diffuse scattering in ray tracing},
  booktitle = {2020 IEEE 31st Annual International Symposium on Personal,
               Indoor and Mobile Radio Communications (PIMRC)},
  year      = {2020},
  pages     = {1--7},
  doi       = {10.1109/PIMRC48278.2020.9217182}
}

@article{HuangTWC2021,
  author  = {Huang, Chen and Wang, Rui and Tang, Pan and He, Ruisi and
             Ai, Bo and Zhong, Zhangdui and Oestges, Claude and
             Molisch, Andreas F.},
  title   = {Geometry-cluster-based stochastic {MIMO} model for
             vehicle-to-vehicle communications in street canyon scenarios},
  journal = {IEEE Transactions on Wireless Communications},
  year    = {2021},
  volume  = {20},
  number  = {2},
  pages   = {755--770},
  doi     = {10.1109/TWC.2020.3028249}
}

@article{KaredalTWC2009,
  author  = {Karedal, Johan and Tufvesson, Fredrik and Czink, Nicolai and
             Paier, Alexander and Dumard, Charlotte and Zemen, Thomas and
             Mecklenbr{\"a}uker, Christoph F. and Molisch, Andreas F.},
  title   = {A geometry-based stochastic {MIMO} model for
             vehicle-to-vehicle communications},
  journal = {IEEE Transactions on Wireless Communications},
  year    = {2009},
  volume  = {8},
  number  = {7},
  pages   = {3646--3657},
  doi     = {10.1109/TWC.2009.080753}
}

@article{SchubertTAP2013,
  author  = {Schubert, Frank M. and Jakobsen, Morten L. and
             Fleury, Bernard H.},
  title   = {Non-stationary propagation model for scattering volumes
             with an application to the rural {LMS} channel},
  journal = {IEEE Transactions on Antennas and Propagation},
  year    = {2013},
  volume  = {61},
  number  = {5},
  pages   = {2817--2828},
  doi     = {10.1109/TAP.2013.2242821}
}

@article{Zemen2025SSCR,
  author  = {Zemen, Thomas and Gomez-Ponce, Jorge and Chandra, Aniruddha and
             Walter, Michael and Aksoy, Enes and He, Ruisi and
             Matolak, David and Kim, Minseok and Takada, Jun-ichi and
             Salous, Sana and Valenzuela, Reinaldo and Molisch, Andreas F.},
  title   = {Site-specific radio channel representation for {5G} and {6G}},
  journal = {IEEE Communications Magazine},
  year    = {2025},
  volume  = {63},
  number  = {6},
  pages   = {106--113},
  doi     = {10.1109/MCOM.001.2400355}
}

@article{Fuschini2015RTReview,
  author  = {Fuschini, Franco and Vitucci, Enrico M. and Barbiroli, Marina and
             Falciasecca, Gabriele and Degli-Esposti, Vittorio},
  title   = {Ray tracing propagation modeling for future small-cell and
             indoor applications: A review of current techniques},
  journal = {Radio Science},
  year    = {2015},
  volume  = {50},
  number  = {6},
  pages   = {469--485},
  doi     = {10.1002/2015RS005659}
}

@article{bernado2014delay,
  author  = {Bernad{\'o}, Laura and Zemen, Thomas and Tufvesson, Fredrik and
             Molisch, Andreas F. and Mecklenbr{\"a}uker, Christoph F.},
  title   = {Delay and {Doppler} spreads of nonstationary vehicular channels
             for safety-relevant scenarios},
  journal = {IEEE Transactions on Vehicular Technology},
  year    = {2014},
  volume  = {63},
  number  = {1},
  pages   = {82--93},
  doi     = {10.1109/TVT.2013.2271956}
}

@inproceedings{Hoydis2023SionnaRT,
  author    = {Hoydis, Jakob and {Ait Aoudia}, Fay{\c{c}}al and
               Cammerer, Sebastian and Nimier-David, Merlin and
               Binder, Nikolaus and Marcus, Guillermo and Keller, Alexander},
  title     = {{Sionna RT}: Differentiable ray tracing for radio
               propagation modeling},
  booktitle = {2023 IEEE Globecom Workshops (GC Wkshps)},
  year      = {2023},
  pages     = {317--321},
  doi       = {10.1109/GCWkshps58843.2023.10465179}
}

@misc{AitAoudia2025SionnaRT,
  author       = {{Ait Aoudia}, Fay{\c{c}}al and Hoydis, Jakob and
                  Nimier-David, Merlin and Nicolet, Baptiste and
                  Cammerer, Sebastian and Keller, Alexander},
  title        = {{Sionna RT}: Technical report},
  year         = {2025},
  howpublished = {arXiv:2504.21719},
  note         = {[Online]. Available: \url{https://arxiv.org/abs/2504.21719}}
}

@article{DegliEsposti2007,
  author  = {Degli-Esposti, Vittorio and Fuschini, Franco and
             Vitucci, Enrico M. and Falciasecca, Gabriele},
  title   = {Measurement and modelling of scattering from buildings},
  journal = {IEEE Transactions on Antennas and Propagation},
  year    = {2007},
  volume  = {55},
  number  = {1},
  pages   = {143--153},
  doi     = {10.1109/TAP.2006.888422}
}

@article{Vitucci2012,
  author  = {Vitucci, Enrico M. and Mani, Francesco and
             Degli-Esposti, Vittorio and Oestges, Claude},
  title   = {Polarimetric properties of diffuse scattering from building
             walls: Experimental parameterization of a ray-tracing model},
  journal = {IEEE Transactions on Antennas and Propagation},
  year    = {2012},
  volume  = {60},
  number  = {6},
  pages   = {2961--2969},
  doi     = {10.1109/TAP.2012.2194683}
}

@article{Slepian1978,
  author  = {Slepian, David},
  title   = {Prolate spheroidal wave functions, {Fourier} analysis,
             and uncertainty---{V}: The discrete case},
  journal = {Bell System Technical Journal},
  year    = {1978},
  volume  = {57},
  number  = {5},
  pages   = {1371--1430},
  doi     = {10.1002/j.1538-7305.1978.tb02104.x}
}

@inproceedings{Zelenbaba2020Sounder,
  author    = {Zelenbaba, Stefan and L{\"o}schenbrand, David and
               Hofer, Markus and Daki{\'c}, Anja and Rainer, Benjamin and
               Zemen, Thomas},
  title     = {A scalable mobile multi-node channel sounder},
  booktitle = {2020 IEEE Wireless Communications and Networking
               Conference (WCNC)},
  year      = {2020},
  pages     = {1--6},
  doi       = {10.1109/WCNC45663.2020.9120769}
}

@inproceedings{Bernado2009,
  author    = {Bernad{\'o}, Laura and Zemen, Thomas and Paier, Andreas and
               Karedal, Johan},
  title     = {Parametrization of the local scattering function estimator
               for vehicular-to-vehicular channels},
  booktitle = {2009 IEEE 70th Vehicular Technology Conference Fall},
  year      = {2009},
  pages     = {1--5},
  doi       = {10.1109/VETECF.2009.5378786}
}

@misc{sionna,
  author       = {Hoydis, Jakob and Cammerer, Sebastian and
                  {Ait Aoudia}, Fay{\c{c}}al and Nimier-David, Merlin and
                  Maggi, Lorenzo and Marcus, Guillermo and Vem, Avinash and
                  Keller, Alexander},
  title        = {{Sionna}},
  year         = {2022},
  howpublished = {[Online]. Available: \url{https://nvlabs.github.io/sionna/}},
  note         = {Software, version 2.2.0}
}

@article{Xia2024,
  author  = {Xia, Guozhen and Zhou, Chen and Zhang, Fubin and Cui, Zeyu and
             Liu, Chengkai and Ji, Hao and Zhang, Xinmiao and
             Zhao, Zhengyu and Xiao, Yin},
  title   = {Path loss prediction in urban environments with {Sionna-RT}
             based on accurate propagation scene models at 2.8 {GHz}},
  journal = {IEEE Transactions on Antennas and Propagation},
  year    = {2024},
  volume  = {72},
  number  = {10},
  pages   = {7986--7997},
  doi     = {10.1109/TAP.2024.3451214}
}

\end{document}